*Research article*

# FCM-DNN: diagnosing coronary artery disease by deep accuracy fuzzy C-means clustering model

**Javad Hassannataj Joloudari[1,*], Hamid Saadatfar[1], Mohammad GhasemiGol[1], Roohallah Alizadehsani[2], Zahra Alizadeh Sani[3,4], Fereshteh Hasanzadeh[4], Edris Hassannataj[5], Danial Sharifrazi[6] and Zulkefli Mansor[7,*]**

[1] Department of Computer Engineering, Faculty of Engineering, University of Birjand, Birjand, Iran
[2] Institute for Intelligent Systems Research and Innovation, Deakin University, Geelong, VIC 3216, Australia
[3] Rajaie Cardiovascular Medical and Research Center, Iran University of Medical Sciences, Tehran, Iran
[4] Omid hospital, Iran University of Medical Sciences, Tehran, Iran
[5] Department of Nursing, School of Nursing and Allied Medical Sciences, Maragheh Faculty of Medical Sciences, Maragheh, Iran
[6] Department of Computer Engineering, School of Technical and Engineering, Shiraz Branch, Islamic Azad University, Shiraz, Iran
[7] Faculty of Information Science and Technology, Universiti Kebangsaan Malaysia, UKM Bangi 43600, Malaysia

* **Correspondence:** Email: javad.hassannataj@birjand.ac.ir, kefflee@ukm.edu.my.

**Abstract:** Cardiovascular disease is one of the most challenging diseases in middle-aged and older people, which causes high mortality. Coronary artery disease (CAD) is known as a common cardiovascular disease. A standard clinical tool for diagnosing CAD is angiography. The main challenges are dangerous side effects and high angiography costs. Today, the development of artificial intelligence-based methods is a valuable achievement for diagnosing disease. Hence, in this paper, artificial intelligence methods such as neural network (NN), deep neural network (DNN), and fuzzy C-means clustering combined with deep neural network (FCM-DNN) are developed for diagnosing CAD on a cardiac magnetic resonance imaging (CMRI) dataset. The original dataset is used in two different approaches. First, the labeled dataset is applied to the NN and DNN to create the NN and DNN models. Second, the labels are removed, and the unlabeled dataset is clustered via the FCM method, and then, the clustered dataset is fed to the DNN to create the FCM-DNN model. By utilizing the second clustering and modeling, the training process is improved, and consequently, the accuracy is increased. As a result, the proposed FCM-DNN model achieves the best performance with a 99.91% accuracy specifying 10 clusters, i.e., 5 clusters for healthy subjects and 5 clusters for sick subjects, through the 10-fold cross-validation technique compared to the NN and DNN models reaching the accuracies of 92.18% and 99.63%, respectively. To the best of our knowledge, no study has been conducted for CAD diagnosis on the CMRI dataset using artificial intelligence methods. The results confirm that the proposed FCM-DNN model can be helpful for scientific and research centers.

## 1. Introduction

Heart disease is an umbrella term that encompasses various diseases, including congenital diseases, CAD, and heart rheumatism. Based on the World Health Organization (WHO) report, CAD is the most common disease in middle-aged and older people, giving rise to killing more than 360,000 Americans in 2015 [1–6]. Moreover, according to the clinical centers for disease control and prevention statistics report, an American experiences a heart attack per 40 seconds [7]. Moreover, more than 75% of deaths have happened due to CAD in developing countries [1]. Regarding the mortality in men and women, more than 50% of the mortality has occurred caused by CAD in men, giving rise to 25% of deaths in the United States [8], and more than 630,000 Americans are dead per year [2], the cost of which has reached more than $ 200 billion [9]. In general, the costs of heart diseases for patients will double by 2030, according to the American Heart Association [10].

Angiography is the most common tool for CAD diagnosis that has side effects and high costs for patients [7]. In scientific centers, researchers use artificial intelligence methods to provide appropriate diagnostic models instead of angiography for CAD diagnosis [11,12]. The methods utilized by artificial intelligence researchers to diagnose CAD are machine learning and deep learning [11,13,14]. In recent years, deep learning (DL) methods have been used for the effective analysis of medical images [15–20].

Accordingly, we propose methods such as neural network (NN), deep neural network (DNN), and fuzzy C-means clustering combined with DNN (FCM-DNN) for CAD diagnosis on cardiac magnetic resonance imaging (CMRI) dataset. The deep neural network (DNN) method is developed as an extended neural network (NN) method, which leads to higher detection accuracy, lower false rate, and lower deviation [21]. In this study, the image set contains labels with healthy and sick classes. To implement, both labeled and unlabeled data are considered for the training process. First, the labeled data is trained and tested using the NN and DNN methods so that the created NN and DNN models are evaluated under the criteria of accuracy, precision, sensitivity, specificity, F1-score, false positive rate, false negative rate, and area under the curve (AUC). Second, since the other model is a hybrid FCM-DNN model, the input data must be unlabeled. For this purpose, the data labels are removed, and the fuzzy C-means clustering method is applied to specify 10 clusters, 5 clusters for healthy subjects, and 5 clusters for sick subjects. Then, the clustered data is fed to the DNN. The created FCM-DNN model is also evaluated under the criteria mentioned above. As a final result, the proposed hybrid FCM-DNN method is a very accurate method with a maximum accuracy of 99.91% compared to the related methods used for CAD diagnosis.

In summary, the innovations of this paper are as follows:

- Providing the CMRI dataset to diagnose CAD for the first time
- According to the latest studies, using the developed FCM-DNN model to diagnose CAD by removing the labels of the data for the first time
- Improving the DNN training and preventing the data over-fitting by performing operations such as selecting Maxout without the need for drop out, using the K-fold Cross-Validation (K-FCV) technique, and feeding the clustered data by the FCM to the DNN
- Achieving a very high accuracy using the proposed hybrid model for diagnosing CAD on the CMRI dataset

As the latest scientific achievement, the FCM-DNN model is performed for the first time using the CMRI analysis.

Currently, tools such as exercise stress testing (EST), chest x-ray, computed tomography scan, CMRI, coronary angiography, and ECG are used to diagnose the severity of heart disease in patients [22,23]. In recent years, more studies have been carried out in the field of CAD diagnosis based on ECG signals and numerical datasets using artificial intelligence methods.

In a study by Babaoglu et al. [24], CAD diagnosis has been made using genetic algorithm (GA), binary particle swarm optimization (BPSO) algorithm, and support vector machine (SVM) algorithm on EST dataset. In addition, GA and BPSO algorithms have been applied as feature selection techniques. In their study, 408 patients have been tested through EST and coronary angiography. A total of 23 features have been extracted from the EST dataset. Using the BPSO algorithm, the diagnosis accuracy rate reaching 81.46% is the best compared to the GA and SVM algorithms achieving 79.17% and 76.67% accuracies, respectively.

Kumar et al. [25] have used ECG signals including 40 healthy subjects and 7 sick subjects for CAD diagnosis. The ECG signals have been mapped into pulses, which were mainly decomposed by analytical wavelet transform. The least-squares support vector machine with the radial basis function (RBF) kernel has been used for classification. As a result, the Violet kernel or Morlet wavelet kernel with the accuracy of 99.60% has provided higher accuracy than the RBF kernel, reaching the accuracy of 99.56% using the 10-fold cross-validation (10-FCV) technique.

Alizadehsani et al. have suggested sequential minimal optimization (SMO) and naïve bayes (NB) classification methods separately and in combination based on ECG symptoms and characteristics for CAD diagnosis on 303 samples [26]. The 10-FCV technique has been used to evaluate the algorithms. As a result, using the SMO-NB algorithm, they have achieved greater accuracy of 88.52% compared to the SMO and NB algorithms with the accuracies of 86.95% and 87.52%, respectively.

In another study, Alizadehsani et al. [27] have proposed classification algorithms such as SMO, NB, bagging with SMO, and neural network for CAD diagnosis on 303 numerical data with 54 features. They have used information gain (IG) and confidence methods to determine the essential features. The SMO algorithm with IG has achieved the best performance with an accuracy rate of 94.08% via the 10-FCV technique.

Alizadehsani et al. have presented C4.5 decision tree and bagging algorithms on a numerical dataset of 303 samples to diagnose CAD disease [28]. They have used IG and gini index (GI) methods for feature selection. In the CAD diagnosis test, coronary artery stenosis has been examined, including left anterior descending (LAD), left circumflex (LCX) (LCX; for the left coronary artery), and Right coronary artery (RCA). In addition, the accuracy of the algorithms has been computed based on the 10-FCV technique. As a result, the bagging algorithm with feature selection method and GI has better performance than the C4.5 decision tree. The accuracy of the bagging algorithm for diagnosing coronary artery stenosis, including LAD, LCX and RCA, has been 79.54%, 65.09% and 66.31%, respectively. However, the C4.5 decision tree has obtained the diagnosis accuracies of 76.56%, 63.10% and 63.38% for stenosis of LAD, LCX and RCA, respectively.

Alizadehsani et al. have applied the SVM method on 303 patients with 54 features to diagnose CAD. The stenosis of the three large coronary arteries has been diagnosed separately based on demographic, symptom and examination, ECG, laboratory, and echo characteristics [29]. They have also used analytical methods to explore the significance of the vascular stenosis features. Using the SVM method with feature selection methods such as combined information gain and average information gain, the accuracy rates of 86.14%, 83.17% and 83.50% have been obtained to diagnose the stenosis of LAD, LCX and RCA, respectively.

Dolatabadi et al. [30] have studied the combined method of principal component analysis (PCA) and optimized SVM to diagnose CAD on heart rate variability (HRV) signal extracted from the ECG. The PCA method has been used to reduce the dimensions of the features and computational complexity. Furthermore, the optimized SVM method refers to optimizing cost and sigma parameters. Therefore, the diagnostic accuracies obtained through the combined method and the standard SVM method have been 99.2% and 90.62%, respectively.

Arabasadi et al. have examined neural network and genetic algorithms for CAD diagnosis on 303 numerical samples [31]. Feature ranking methods such as weight by SVM, GI, IG and PCA have been applied for feature selection. As a result, using the neural network algorithm and the combined neural network-genetic algorithm, the accuracy rates of 84.62% and 93.85% have been obtained, respectively.

Alizadehsani et al. [32] have utilized a feature engineering method for improving CAD diagnosis on the 500 samples. This method has exploited the results related to the NB, C4.5, and SVM classifiers for the non-invasive diagnosis of CAD disease. They have also used the weight by SVM method as a feature selection method. Based on the NB, C4.5, and SVM classifiers, the accuracy rates of 86%, 89.8% and 96.40% have been achieved, respectively.

In a study by Abdar et al. [33], a hybrid two-level genetic algorithm and nuSVM, namely the N2Genetic-NuSVM method, has been proposed for CAD diagnosis on 303 samples. They have used a two-level genetic algorithm to optimize the SVM parameters and also have accomplished feature selection applying the GA algorithm. An accuracy of 93.08% has been achieved using the proposed method.

In the study conducted by Miao and Miao [34], a DNN model has been presented for CAD diagnosis on the Cleveland Clinic Foundation dataset with 303 patients. The proposed DL model includes 28 input units, first and second hidden layers, and a binary output unit, in which 105 neurons in the first layer and 42 neurons in the second layer have been considered, and 50% dropout has been assigned. The output unit has been connected to a sigmoid activation function in the final stage. An accuracy of 83.67% has been obtained using the proposed method.

Hamersvelt et al. have examined a convolutional neural network (CNN) to diagnose CAD on the coronary artery angiography CT images at rest with 126 patients [35]. As a result, by applying the proposed CNN method, an accuracy of 71.7% was achieved.

Hassannataj et al. [17] have extracted essential CAD features, which was diagnosed on 303 samples with 55 features using the random trees (RTs). They have compared the RTs model with the SVM, the C5.0 decision tree, and the CHAID decision tree classification models. As a result, the RTs model has provided the best performance compared to the other models by extracting 40 features with an accuracy of 91.47%.

Acharya et al. [36] have implemented the CNN method for CAD diagnosis, applying different periods of ECG signal segments from the PhysioNet database [37]. In their study, an 11-layer CNN structure, including four convolutional layers, four max-pooling layers, and three fully-connected layers, has been developed. Moreover, an overall of 95,300 segmented ECG signals have been utilized for the first network (2 seconds), and a total of 38,120 segmented ECG signals have been used for the second network (5 seconds). The proposed CNN model has achieved the accuracies of 94.95% and 95.11% for the first and second networks, respectively.

Tan et al. [38] have introduced a long short-term memory (LSTM) neural network model combined with a CNN model for CAD diagnosis based on ECG signals on the PhysioNet database. Accordingly, an 8-layer stacked convolutional LSTM network has been designed, in which layers 1 to 4 consist of two convolutional layers and two layers of max pooling for the CNN structure, layers 5 to 7 relate to the LSTM layers, and the last layer is a fully connected layer as the classification layer. The proposed method has achieved an accuracy of 99.85%.

Acharya et al. [39] have investigated the K-nearest neighbor (KNN) classifier to classify and diagnose CAD on ECG signals. To extract features, they have used methods such as discrete cosine transform, discrete wavelet transform, and empirical signal decomposition into intrinsic state components. Besides, these methods have been compared in the disease diagnosis process. The ECG signals have also been applied to the appropriate transformation methods to obtain coefficients. Then, the features have been reduced using the locality preserving projection method, and the reduced features have been ranked applying the analysis of variance technique. In the following, high-ranking features have been fed to the KNN classifier. As a result, the proposed model provided the best performance reaching the accuracy of 98.5% via the discrete cosine transform method using only seven features.

Acharya et al. [40] have presented a CNN method to diagnose CAD on ECG signals. The dataset includes 30,000 patients and 110,000 healthy persons. As a result, the proposed method leads to an accuracy of 98.97% using the 10-FCV technique.

In a study by Ghiasi et al. [41], a regression and classification tree model under the CART model has been investigated on the Z-Alizadeh Sani dataset [27] with 303 patients and 55 features to diagnose CAD. They have compared their model with classification models such as SMO, bagging, bagging with SMO, NB, artificial neural network, and J48 and C4.5 decision trees. The accuracy rate of 100% has been gained using the CART model for CAD diagnosis.

To identify the risk factors for CAD, Verma et al. [42] have implemented a combined model of correlation-based feature subset (CFS) selection with particle swarm optimization (PSO) and K-Means clustering algorithms on 335 samples with 26 features. After applying CFS and PSO, five features have been identified as risk factors. In addition, multi-layer perceptron (MLP), multinomial logistic regression (MLR), fuzzy unordered rule induction algorithm, and C4.5 decision tree have been implemented for CAD diagnosis. As a result, the highest accuracy of 88.4% has been obtained using the MLR algorithm.

Idris et al. [43] have developed data mining models, including NN, logistic regression (LR), KNN, NB, SVM, deep learning, and Vote (an ensemble method with NB and LR) on the Malaysian National Cardiovascular Disease-Acute Coronary Syndrome datasets from the University of Malaya medical centre (UMMC) and Sultanah Aminah hospital (SAH) to predict the CAD. Feature selection methods such as the Chi-squared test, recursive feature elimination, and embedded decision tree have been applied. The prediction accuracy rates of 94.5% and 89.7% have been obtained through the NN method combined with the embedded decision tree method on the UMMC and SAH datasets, respectively.

Velusamy and Ramasamy [44] have examined three classification methods, including SVM, random forest, and KNN, for CAD diagnosis on the Z-Alizadeh Sani dataset. The results of the classifiers have been combined based on the weighted-average voting, majority-voting, and average-voting methods. The weighted-average voting method and five selected features lead to better performance with an accuracy rate of 98.97% compared to other classifiers on the original Z-Alizadeh Sani dataset. In addition, the proposed algorithm reaches the accuracy of 100% on the Z-Alizadeh Sani balanced dataset.

According to the previous works, researchers have investigated three types of datasets, including numerical, CT scan, and ECG signal datasets for CAD diagnosis. In this paper, we have utilized the MRI dataset to diagnose CAD for the first time. The strength of this research is the use of the CMRI dataset in two labeled and unlabeled forms, with the NN and DNN methods applied to the labeled data and the FCM-DNN method applied to the unlabeled data. Moreover, in the previous works, the important accuracy evaluation criterion has been calculated on labeled data, while in our paper, a great accuracy rate of 99.91% has been obtained based on the FCM method in combination with the DNN classifier on the unlabeled data.

The rest of the paper is structured as follows: The proposed methodology is introduced in Section 2. The evaluation of models, experimental results, and research findings are expressed in Section 3. Comparison with the previous researches and discussion are presented in Section 4. Finally, the conclusion and future work are given in Section 5.

## 2. Methodology

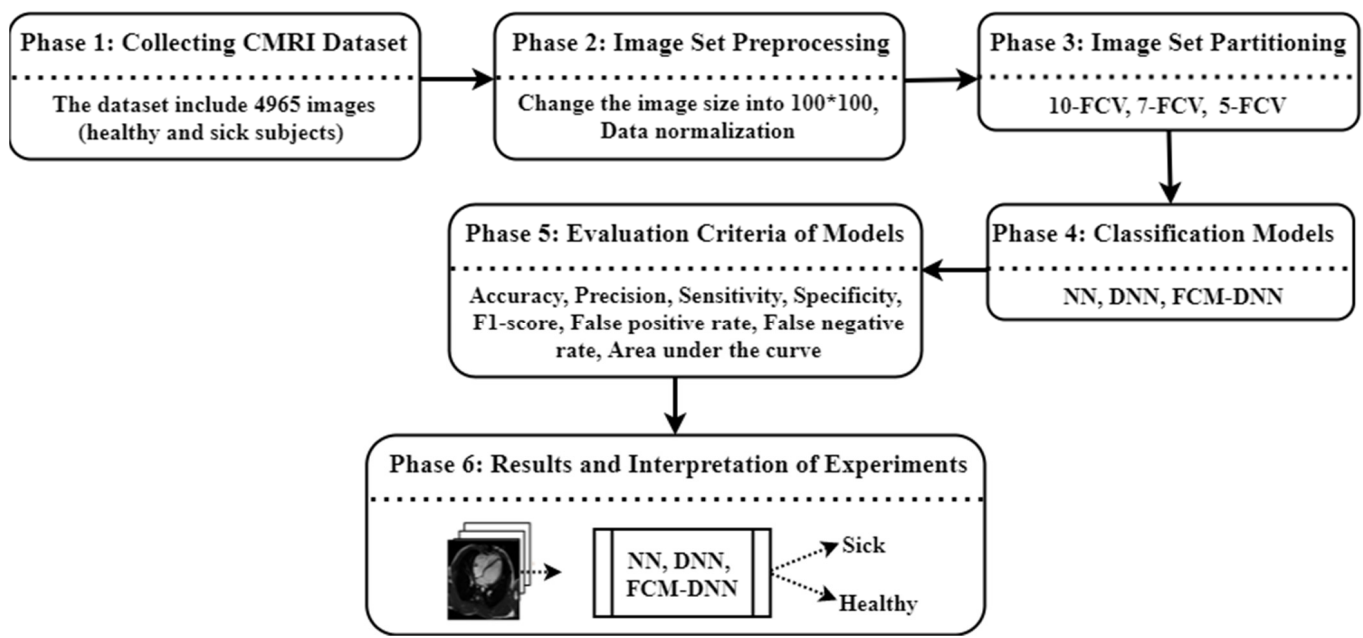

**Figure 1.** The proposed methodology.

In this paper, NN, DNN and fuzzy C-means clustering combined with deep neural network (FCM-

DNN) methods are used on the CMRI dataset to classify the images and diagnose the CAD. The proposed methodology is implemented in 6 phases, including collecting clinical image sets related to CMRI dataset for healthy and sick subjects, data preprocessing, CMRI dataset partitioning, classification models, evaluation criteria of the models, experimental results, and their interpretation for classification of the CMRI images and diagnosis of the CAD. The proposed methodology is shown in Figure 1.

## 2.1. Phase 1: collecting clinical CMRI dataset

The first phase is the extraction of CAD clinical image sets related to CMRI. This dataset is provided from Milad hospital in Tehran, Iran, by Z. Alizadeh Sani. The dataset utilized in this paper includes 4965 images so that 2569 images of which are related to 16 healthy subjects, and the remaining 2396 images are associated with 14 sick subjects. All the images are grayscale, and their dimensions vary for healthy and sick subjects. For example, four images of healthy and sick subjects are illustrated in Figures 2 and 3, respectively.

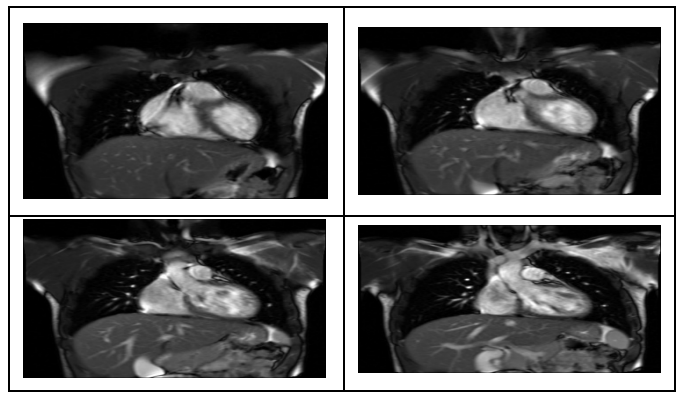

**Figure 2.** The images of healthy subjects.

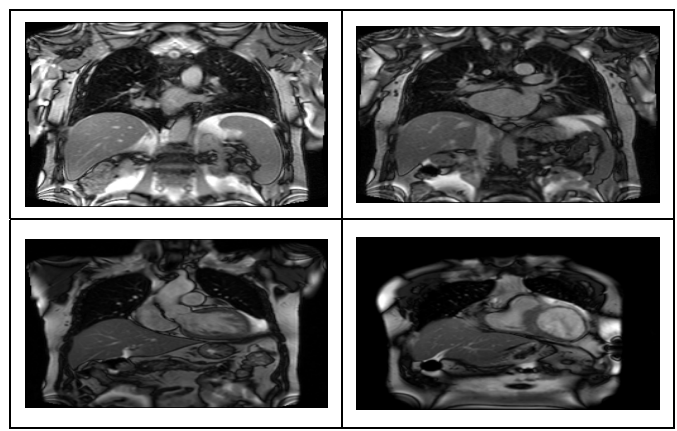

**Figure 3.** The images of sick subjects.

In addition, the statistical characteristics of our dataset for healthy and sick subjects are stated in Tables 1 and 2, respectively.

**Table 1.** Statistical characteristics of the CMRI dataset for healthy subjects.

| Statistics | Age | Weight | Height |
|---|---|---|---|
| Mean | 34.75 | 65.13 | 165.75 |
| Std. Error of Mean | 4.955 | 4.531 | 3.322 |
| Median | 34.50 | 69.50 | 170.00 |
| Std. Deviation | 19.821 | 18.125 | 13.289 |
| Variance | 392.867 | 328.517 | 176.600 |

**Table 2.** Statistical characteristics of the CMRI dataset for sick subjects.

| Statistics | Age | Weight | Height |
|---|---|---|---|
| Mean | 59.86 | 70.79 | 170.71 |
| Std. Error of Mean | 3.797 | 2.881 | 2.286 |
| Median | 64.50 | 72.00 | 169.50 |
| Std. Deviation | 14.206 | 10.779 | 8.552 |
| Variance | 201.824 | 116.181 | 73.143 |

## 2.2. Phase 2: data preprocessing

In the samples analysis process, preprocessing the samples is required. The images of healthy and sick subjects in the dataset differ in size, thus their size is changed into a $100 \times 100$ dimension. Furthermore, one of the available approaches for preprocessing image samples is data normalization between 0 and 1. Normalization increases the accuracy of clustering and classification models and also reduces the false rate of clustering. The type of normalization method is determined as interval transformation, i.e., the sample set is normalized between 0 and 1. Indeed, by normalizing the images, the light intensity of the images is in the interval of 0 and 1.

## 2.3. Phase 3: image set partitioning

In this paper, the K-FCV technique is applied for the partitioning phase of the CMRI dataset, i.e., the data is divided based on the 10-FCV, 7-FCV, and 5-FCV techniques. Utilizing the K-FCV technique, the images are divided into K parts so that K-1 parts are used for training and 1 part for testing. By rotating the test image set, the K-FCV process is repeated K times. The advantages of the K-FCV technique are that this technique prevents data over-fitting, improves training, and reduces loss. Moreover, applying the K-FCV technique leads to more training data points to develop the expected model.

Therefore, the dataset is partitioned based on the 10-FCV, 7-FCV, and 5-FCV techniques. The partitioning process for training, testing, and validating the CMRI dataset through the 10-FCV, 7-FCV, and 5-FCV techniques is shown in Figures 4, 5 and 6, respectively.

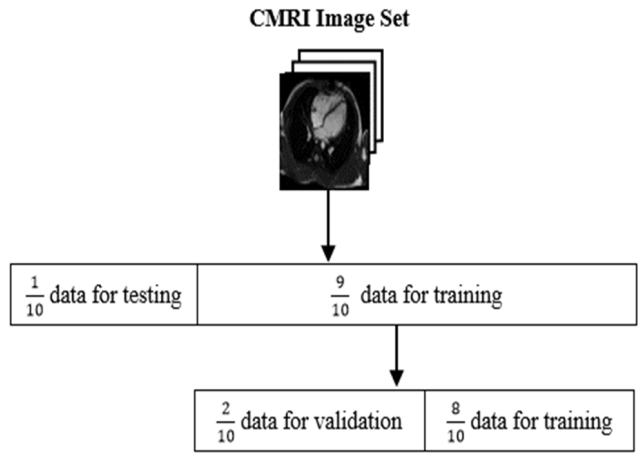

**Figure 4.** The partitioning process using the 10-FCV technique.

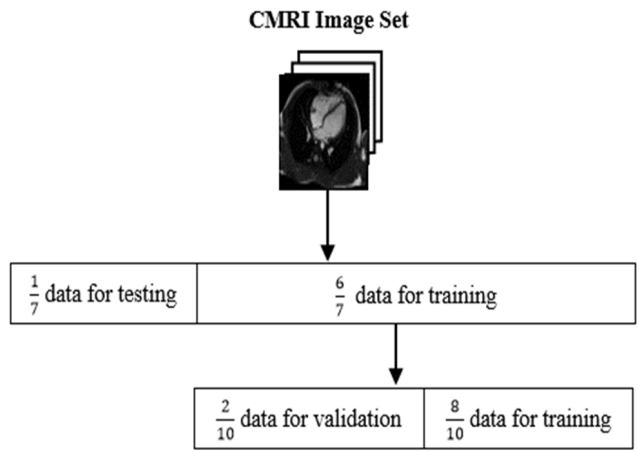

**Figure 5.** The partitioning process using the 7-FCV technique.

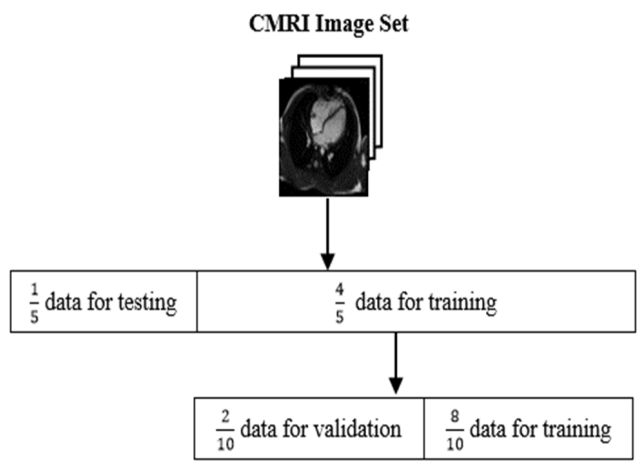

**Figure 6.** The partitioning process using the 5-FCV technique.

According to Figures 4–6, the 10-FCV, 7-FCV and 5-FCV techniques are utilized for training the models, respectively. 0.9 of the data is used for training, and the remaining 0.2 is utilized for testing via the 10-FCV technique. Six-seventh of the data is applied for training, and the remaining one-seventh is used for testing through the 7-FVC technique. Four-fifth of the data is utilized for training, and the remaining one-fifth is applied for testing by the 5-FCV technique. In the next step, considering 0.8 of the training data for training and 0.2 of the training data for validation, the partitioning process is accomplished 10, 7 and 5 times for 10-FVC, 7-FCV and 5-FCV techniques, respectively.

### 2.4. Phase 4: classification models

The most common system for diagnosing CAD is angiography. This system has many side effects and high costs for patients. On the other hand, CAD can lead to myocardial infarction if the patient's condition is not correctly diagnosed during testing and also is not treated on time. Therefore, it is essential to use intelligent automated decision-making systems and technologies for CAD diagnosis. In recent years, researchers have tried to use artificial intelligence techniques as an alternative to angiography for the early diagnosis of CAD. Hence, in this paper, NN, DNN, and FCM-DNN classification methods are proposed to be applied to the CMRI dataset. The creation of the NN, DNN and FCM-DNN models is described in detail below.

#### 2.4.1. Creating NN model

The structure of the NN is derived from the structure of the human neural network in the biological brain. In the human neural network, there are a series of functional units called cells and neurons. In neurons, the data is in the form of pulses that enter and exit the cell so that as the pulse passes through the cell, a series of processes take place in the cell nucleus. This process is learned all over human life, and the so-called neural network structure is trained throughout human life.

**Table 3.** Parameters settings for the NN model.

| Parameters | Settings |
|---|---|
| Training cycles | 20 |
| Momentum | 0.2 |
| Number of hidden layers | 2 |
| Hidden layer size | $50 \times 50$ |
| Shuffle | ✓ |
| Normalize | ✓ |
| Epsilon | 0.001 |
| Shrinking | ✓ |

In general, the standard NN is one of the classification methods in which the created model is identified as a set of interconnected nodes with their weighted connections. This created model includes the input layer, hidden layer, and output layer. The process of generating output is such that each of the input dimensions is multiplied by a weight factor. Then, the sum of the multiplications of these weights passes through a nonlinear function, which eventually produces a new output. In other words, in this neural network, there is a layer called the feature layer or hidden layer, the output of which is the feature space, and the input to the last layer, i.e., the classifier layer, which determines the class of the input data.

Hence, in this paper, the created NN model is a 4-layer model. The first layer is related to the input images. The second and third layers are specified as the hidden layers with three neurons, and the last layer is determined as output class (sick/healthy). The parameters settings for the NN model are described in Table 3, and the NN model is presented in Figure 7.

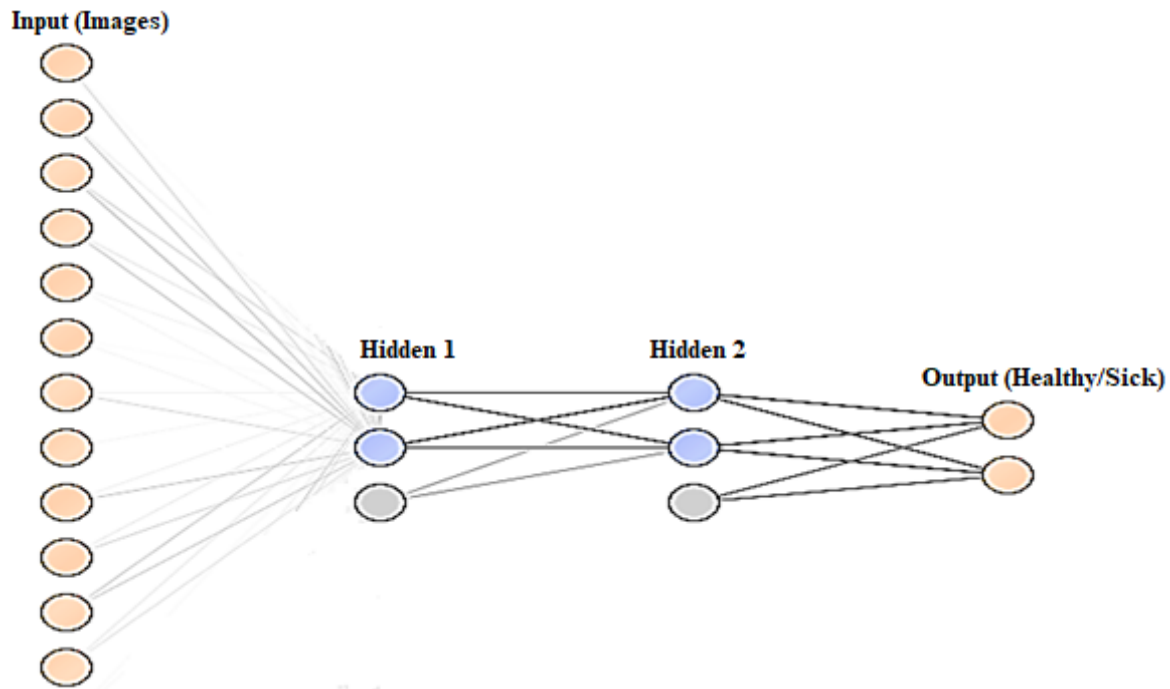

**Figure 7.** The NN model.

The pseudo-code of the NN model is presented below.

| | Algorithm 1. NN model for CAD diagnosis |
| --- | --- |
| | **Input**: The Z-Alizadeh Sani image set including 4965 images |
| | **Output**: The diagnosed Sick/Healthy class for each test image and obtained evaluation criteria for the created model |
| 1. | **Begin** |
| 2. | Image set preprocessing: Change the image size to $100 \times 100$ and normalize the data |
| 3. | Divide the data using 10-FCV, 7-FCV, and 5-FCV techniques |
| 4. | **While** (The termination condition is not fulfilled by 10-FCV, 7-FCV, and 5-FCV techniques or considering the number of training cycles) do |
| 5. | Apply NN training for each image |
| 6. | Create NN model |
| 7. | Apply NN validation for each image |
| 8. | Apply NN model for testing input images |
| 9. | **End while** |
| 10. | **Return** Obtain the evaluation criteria and diagnose the Sick/Healthy classes for input images |
| 11. | **End** |

### 2.4.2. Creating DNN model

The standard NN model is such that it tends to have a high error deviation, which can lead to adverse effects on the classification performance. To address this problem, the DNN model, as an extended NN model, improves the classification performance by increasing the number of hidden layers from 3 to 100 and more. Another strength of the DNN model is that in this model, the data is transferred from one hidden layer to another so that simpler features are recombined and recomposed into complex features to generate the desired output. The advantages of the deep learning model are expressed below:

1) Automated feature learning: The DNN model automatically extracts appropriate features from the data and is so-called trained.

2) Multi-layer feature learning: Based on the DNN model, there is the ability to simultaneously access features at different levels in a hierarchical manner, from low-level features to complex level features.

3) High accuracy of deep neural network diagnosis: The accuracy of the DNN model in the output is higher than the accuracy of the NN model.

4) High generalization power of the network: High generalization power means that in addition to the data trained by the DNN if new data similar to the training data is fed to the network, the highly developed DNN model can diagnose the new data as well.

In the DNN model, similar to the NN model, the images are applied to the input layer, and the class of the input images is specified in the output layer.

**Table 4.** Parameters settings for the DNN model.

| Activation function | Maxout |
| --- | --- |
| Epochs | 50 |
| Number of hidden layers | 6 |
| Hidden layer size | $50 \times 40 \times 30 \times 20 \times 15 \times 10$ |
| Training samples per iteration | $-2$ (auto-tuning) |
| Adaptive rate | ✓ |
| Epsilon | 1.0E−8 |
| Rho (similar to momentum and relevant to the memory to former weight updates) | 0.99 |
| Standardize | ✓ |
| L1- regularization | 1.0E−5 |
| L2- regularization | 0.0 |
| Loss function | Cross-Entropy |
| Distribution function | Bernoulli |
| Activation function in the last layer | Sigmoid function |

Therefore, in this paper, the created DNN model is an 8-layer model including one input layer for the images, six hidden layers, and one output layer. The Maxout [45,46] is selected as a nonlinear activation function, which assigns the activity of the neurons to the hidden layers of the network. Indeed, the Maxout determines the utmost coordinate of the network input vector, which is effective for over-fitting of the input data, reducing the complexity, and improving the deep network training.

The Maxout function is defined as follows for two classes:

$$g_i(x) = Max(g_1(x) = w_0 + w_{11}.x_1 + w_{12}.x_2, \; g_2(x) = w_0 + w_{21}.x_1 + w_{22}.x_2) \tag{1}$$

According to Eq (1), $w_0$, $w_{ij}$, and $x_i$ represent the random initial weight, the elements entries of the weights, and the feature vector of the input images for the sick and healthy classes, respectively.

The parameters settings for the DNN model are described in Table 4, and the DNN model is presented in Figure 8.

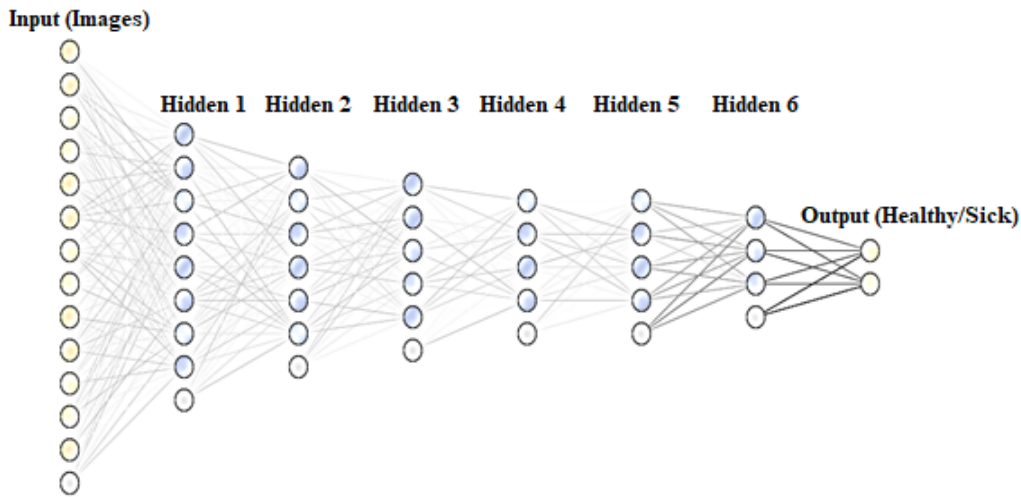

**Figure 8.** The DNN model.

To classify the healthy and sick subjects by determining the value of 1 for the healthy subject and the value of 0 for the sick subject, a sigmoid function [47,48] is assigned to the last layer. Moreover, a cross-entropy (CE) function [49] is defined as the loss function. The formulas of these functions are defined as follows:

$$F(S_i) = \frac{1}{1+e^{-S_i}}, \; S = F(x_i, w) \tag{2}$$

$$CE(S.y) = -\sum_{i=1}^{c} y_i \log\big(F(S_i)\big) \tag{3}$$

In Eq (2), the output value of the decision boundary (*Si*) or the probability value of the predicted class is 0 or 1, $x_i$ is the input image, and *w* is the weight. In Eq (3), *C* represents the number of classes, and $y_i$ indicates the predicted value of the desired class. Since, in this paper, the number of classes is two, the *CE* function is calculated as below:

$$CE(S.y) = -\sum_{i=1}^{c} y_i \log\big(F(S_i)\big) = -y_1 \log\big(F(S_1) - (1-y_1)\log(1-F(S_1)\big) \tag{4}$$

According to Eq (4), *F(S₂)* is equal to *1-F(S₁)*.

Also, The pseudo-code of the DNN model is presented below.

| | **Algorithm 2.** DNN model for CAD diagnosis |
|---|---|
| | **Input**: The Z-Alizadeh Sani image set including 4965 images |
| | **Output**: The diagnosed Sick/Healthy class for each test image and obtained evaluation criteria for the created model |
| 1. | **Begin** |
| 2. | Image set preprocessing: Change the image size to $100 \times 100$ and normalize the data |
| 3. | Divide the data using 10-FCV, 7-FCV, and 5-FCV techniques |
| 4. | **While** (The termination condition is not fulfilled by 10-FCV, 7-FCV, and 5-FCV techniques or considering the number of epochs) do |
| 5. | Train samples per iteration based on Table 4 |
| 6. | Apply DNN training for each image |
| 7. | Create DNN model |
| 8. | Apply DNN validation for each image |
| 9. | Apply DNN model for testing input images |
| 10. | Assign sigmoid function (Eq (2)) for classifying Sick/Healthy subjects in the output layer |
| 11. | Calculate loss function through Eq (4) |
| 12. | **End while** |
| 13. | **Return** Obtain the evaluation criteria and diagnose the Sick/Healthy classes for input images |
| 14. | **End** |

### 2.4.3. Creating FCM-DNN model

Clustering is a standard descriptive method identifying a finite set of categories/clusters for describing similar data. In other words, clustering is the grouping of samples with similar characteristics. The samples of one group have the most similarity to each other and the most difference from the samples of other groups. Each cluster has a center that the degree of the similarity of the data to the center of the cluster is generally determined by a parameter called the similarity criterion/distance criterion [50]. Indeed, in clustering, the similarity criterion is determined based on maximizing the separation between clusters.

Therefore, in clustering, the categories are not predefined, and the data grouping operation is done without supervising or labeling, i.e., the training data do not have a label. The suitable performance of a clustering method is such that the samples of different clusters have the least similarity.

A standard clustering model is shown in Figure 9.

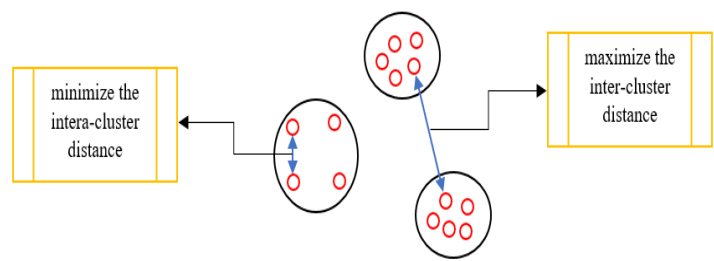

**Figure 9.** The concept of clustering.

In general, in all the clustering methods, the goal is to minimize the intera-cluster distance and maximize the inter-cluster distance [51].

Standard clustering algorithms in vector quantification are K-Means [52,53], K-Medoids [54], and FCM. In this paper, the aim is to create the model in two ways. First, the labeled dataset is fed to the NN and DNN to create the NN and DNN models of the data. Second, the labels are removed from the initial dataset, and the unlabeled data is clustered using the FCM method, and then, the clustered data is fed to the DNN to create the hybrid FCM-DNN model. Here, the utilized FCM clustering method is explained in detail.

The FCM method was first proposed in 1973 by Duda and Hart [55], which performs a more accurate clustering than classic clustering methods, such as K-Means and K-Medoids, under uncertain conditions. Unlike classical clustering methods, fuzzy clustering methods are appropriate for allocating data to more than one cluster.

A fuzzy version of the C-Means clustering method has been proposed by Don to solve the problem of allocating images to more than one cluster [56]. Later, the FCM method was developed by Bezdek [57], in which a fuzzy factor of $m$ has been defined as a fuzzifier.

The main idea of the FCM clustering is that a sample can belong to more than one cluster with a membership degree between 0 and 1 based on the membership function/objective function [58–61].

In the FCM method, the membership function is as follows:

$$\min j_m(u.v) = \Sigma_{i=1}^{c} \Sigma_{k=1}^{n} u_{ik}^{m} d_{ik}^{2} = \Sigma_{i=1}^{c} \Sigma_{k=1}^{n} u_{ik}^{m} \left\| x_k - v_i \right\|^2 \tag{5}$$

where the variable $m$ is a real number larger than 1, which is assigned to be 2 in most cases. In the given formula, if the variable $m$ is set equal to 1, the objective function of C-Means clustering will be obtained. Moreover, in the stated formula, the variable $X_k$ is the sample $K$, $V_i$ is the cluster center, the number of clusters "$C$" is predetermined, and $n$ represents the number of samples. $U_{ik}$ indicates the degree of belonging the sample $i$ to the cluster $k$.

The distance between the sample $X_k$ from the cluster center $V_i$ is computed as follows:

$$d_{ik} = \left\| x_k - v_i \right\| \tag{6}$$

The most crucial similarity criterion for solving clustering problems is the distance criterion "$d$", which must be minimized. In other words, the FCM method determines the data for each cluster based on the distance between the cluster center and the data points by assigning membership to each data based on the membership function.

In summary, the FCM method includes the following steps:

- $C$ cluster centers are randomly assigned.
- The distance of each sample from the center of the cluster is obtained as:

$$U_{imi} = \frac{\frac{1}{d_{imi}}}{\frac{1}{d_{imi}} + \frac{1}{d_{imj}}} \tag{7}$$

where $d$ represents the distance between each sample from the cluster centers $m_i$ and $m_j$, and $U_{imi}$ indicates the degree of belonging to each sample.

- New centers of the clusters are obtained using fuzzy means. If we have two clusters, the new centers of the clusters are achieved as follows:

$$m_1 = \frac{\sum_{i=1}^{N} X_i U_{i1}}{\sum_{i=1}^{N} U_{i1}}, \; m_2 = \frac{\sum_{i=1}^{N} X_i U_{i2}}{\sum_{i=1}^{N} U_{i2}} \tag{8}$$

where $X_i$ represents the sample $i$, and $m_1$ and $m_2$ are the cluster centers.

- Finally, the fuzzy intra-cluster-based sum of the distances is calculated under the following membership function "$J$", which must be optimized:

$$J = \sum_{j=1}^{C} \sum_{i=1}^{N} U_{ij} d_{ij} \tag{9}$$

Based on Eq (8), $J$ is the sum of the distances, "$C$" is the number of clusters, "$U_{ij}$" is the degree of belonging the sample $i$ to the cluster $j$, and "$d_{ij}$" is the distance of sample $i$ from the center $j$. For two consecutive iterations, if the sum of the distances is less than the threshold value, the FCM method will terminate. In this situation, new cluster centers will be determined.

The parameters settings for the FCM method are presented in Table 5.

**Table 5.** Parameters settings for the FCM method.

| Parameters | Settings |
|---|---|
| Add cluster attribute | ✓ |
| Add partition matrix | ✓ |
| Iterations | 50 |
| Fuzzynes | 2.0 |
| MinGain | 1.0E–4 |
| Measure type | MixedMeasures |
| Mixed measure | MixedEuclideanDistance |

Therefore, the advantage of the proposed FCM method is that this method is always convergent and always has a rapid convergence rate in reaching the final solution, i.e., the FCM method converges to a local optimum.

Despite the advantages of the FCM method, the disadvantage of this method compared to the classic clustering methods is its more computational time due to additional calculations for allocating each data to all the clusters. However, the crucial advantage of data clustering using the FCM method is achieving higher accuracy.

In this paper, the FCM-DNN method is examined on the CMRI dataset. Firstly, the dataset is clustered for identifying the clusters by the FCM method. The number of the clusters is assigned as 10. It should be noted that the images were initially labeled as healthy and sick subjects. The labels have been removed for clustering. Then, 10 clusters are determined for clustering operations; 5 clusters for healthy subjects and 5 clusters for sick subjects.

After applying fuzzy clustering, the generated dataset is fed to the DNN model for classifying the CMRI dataset. The developed FCM-DNN model diagnoses the input image between 10 clusters, including healthy and sick classes. The pseudo-code of the FCM-DNN model is presented below.

| | |
|---|---|
| **Algorithm 3.** Hybrid FCM-DNN model for CAD diagnosis | |

**Input**: The Z-Alizadeh Sani image set including 4965 images

**Output**: The diagnosed Sick/Healthy class for each test image and obtained evaluation criteria for the created model

1. **Begin**
2.    Image set preprocessing: Change the image size to $100 \times 100$ and normalize the data
3.    Divide the data using 10-FCV, 7-FCV, and 5-FCV techniques
4.    **While** (The termination condition is not fulfilled by 10-FCV, 7-FCV, and 5-FCV techniques or the iterations are less than 50 based on Table 5) do
5.       Choose $C$ cluster centers randomly ($C = 10$)
6.       Compute the distance of each sample to $C$ cluster centers and determine the degree of belonging to each sample based on Eqs (5) to (7)
7.       Obtain new centers for the clusters using fuzzy means based on Eq (8)
8.       Calculate the fuzzy intra-cluster-based sum of the distances based on Eq (9)
9.       Generate the dataset based on 10 clusters (5 clusters for healthy subjects and 5 clusters for sick subjects)
10.       Apply DNN training for each image according to the generated dataset
11.       Create FCM-DNN model
12.       Apply FCM-DNN validation for each image
13.       Apply FCM-DNN model classification for testing input images
14.       Assign sigmoid function (Eq (2)) for classifying Sick/Healthy subjects in the output layer
15.       Compute loss function
16.    **End while**
17.    **Return** Obtain the evaluation criteria and diagnose the Sick/Healthy classes for input images
18. **End**

## 3. Evaluation and experimental results

In this section, we have evaluated the models based on the fifth phase of the proposed methodology. The evaluation criteria of the models, including accuracy (ACC), precision or positive predicted value (PPV), sensitivity (SEN), specificity (SPC), F1-score, false positive rate (FPR), false negative rate (FNR), and area under the curve (AUC), are measured using a confusion matrix (CM) [62]. The CM includes true positive (TP), false positive (FP), true negative (TN), and false negative (FN) elements. The CM utilized in this paper is described in Table 6.

**Table 6.** The CM for this paper.

| Actual class | The predicted class | |
|---|---|---|
| | Sick (Positive) | Healthy (Negative) |
| Positive | TP | FP |
| Negative | FN | TN |

According to Table 6, the elements of the CM are defined as follows:

TP: The number of positive samples correctly diagnosed as patients by testing.

FP: The number of positive samples wrongly diagnosed as healthy by testing.

TN: The number of negative samples correctly diagnosed as healthy by testing.

FN: The number of negative samples wrongly diagnosed as patients by testing.

Indeed, the CM is a valuable tool for analyzing how the classification models diagnose the data in different categories. If the data is in the $M$ category, a classification matrix will be a table with a minimum size of $M \times M$. Ideally, the TP and TN elements on the main diagonal of the matrix have the highest values, and the rest of the matrix elements have values equal to zero or close to zero [63].

The formulas of the evaluation criteria for the models are given below:

$$ACC = \frac{TP+TN}{FP+FN+TP+TN} \tag{10}$$

$$PPV = \frac{TP}{TP+FP} \tag{11}$$

$$SPC = \frac{TN}{TN+FP} \tag{12}$$

$$SEN = \frac{TP}{TP+FN} \tag{13}$$

$$F1-Score = \frac{2TP}{2TP+FP+FN} \tag{14}$$

$$FPR = 1 - SPC \tag{15}$$

$$FPR = 1 - SEN \tag{16}$$

**Table 7.** Results of the models' evaluation criteria on the CMRI dataset.

| Models | Number of Folds | ACC (%) | PPV (%) | SEN (%) | SPC (%) | F1-Score (%) | FPR | FNR | AUC (%) |
|--------|-----------------|---------|---------|---------|---------|--------------|-----|-----|---------|
| NN | 5 | 92.11 | 89.66 | 95.57 | 88.55 | 92.5 | 11.45 | 4.43 | 93.4 |
| | 7 | 91.79 | 89.89 | 94.57 | 88.93 | 92.07 | 11.07 | 5.43 | 93.6 |
| | 10 | 92.18 | 89.28 | 96.44 | 87.79 | 92.66 | 12.21 | 3.56 | 93.3 |
| DNN | 5 | 99.35 | 99.19 | 99.54 | 99.15 | 99.36 | 0.85 | 0.46 | 99.9 |
| | 7 | 99.44 | 99.72 | 99.18 | 99.72 | 99.45 | 0.28 | 0.82 | 100 |
| | 10 | 99.63 | 99.59 | 99.68 | 99.58 | 99.64 | 0.42 | 0.32 | 99.9 |
| **FCM-** | 5 | 99.66 | 100 | 99.31 | 100 | 99.65 | 0 | 0.69 | 100 |
| **DNN** | 7 | 99.70 | 99.86 | 99.54 | 99.86 | 99.7 | 0.14 | 0.46 | 100 |
| | **10** | **99.91** | **100** | **99.82** | **100** | **99.91** | **0** | **0.18** | **100** |

It should be noted that the FPR criterion is more important than the FNR criterion for clinical centers in identifying more risks in sick subjects [63,64].

According to the sixth phase of the proposed methodology, the experimental results of the models are illustrated in Table 7 in terms of the evaluation criteria and the number of folds. The experimental environment includes Intel(R) Core(TM) i5-4200U CPU @ 1.60 GHz to 2.30 GHz, 6 GB of RAM,

Windows 10 operating system, x64-based processor, and NVIDIA GeForce840M, and the methods are implemented using the RapidMiner software version 9.5.0[1] [46].

According to Table 7, the ACC, PPV, SEN, SPC, F1-score, FPR, FNR and AUC rates are obtained using the NN, DNN and FCM-DNN methods on the CMRI dataset.

The most crucial criterion for diseases diagnosis is ACC. The ACC rate for CAD diagnosis using the proposed FCM-DNN method is more than the NN and DNN methods.

The ACC of the FCM-DNN method is obtained as 99.91% on 4965 images using the 10-FCV technique, while the accuracy of the DNN and NN methods is achieved as 99.63% and 92.18%, respectively.

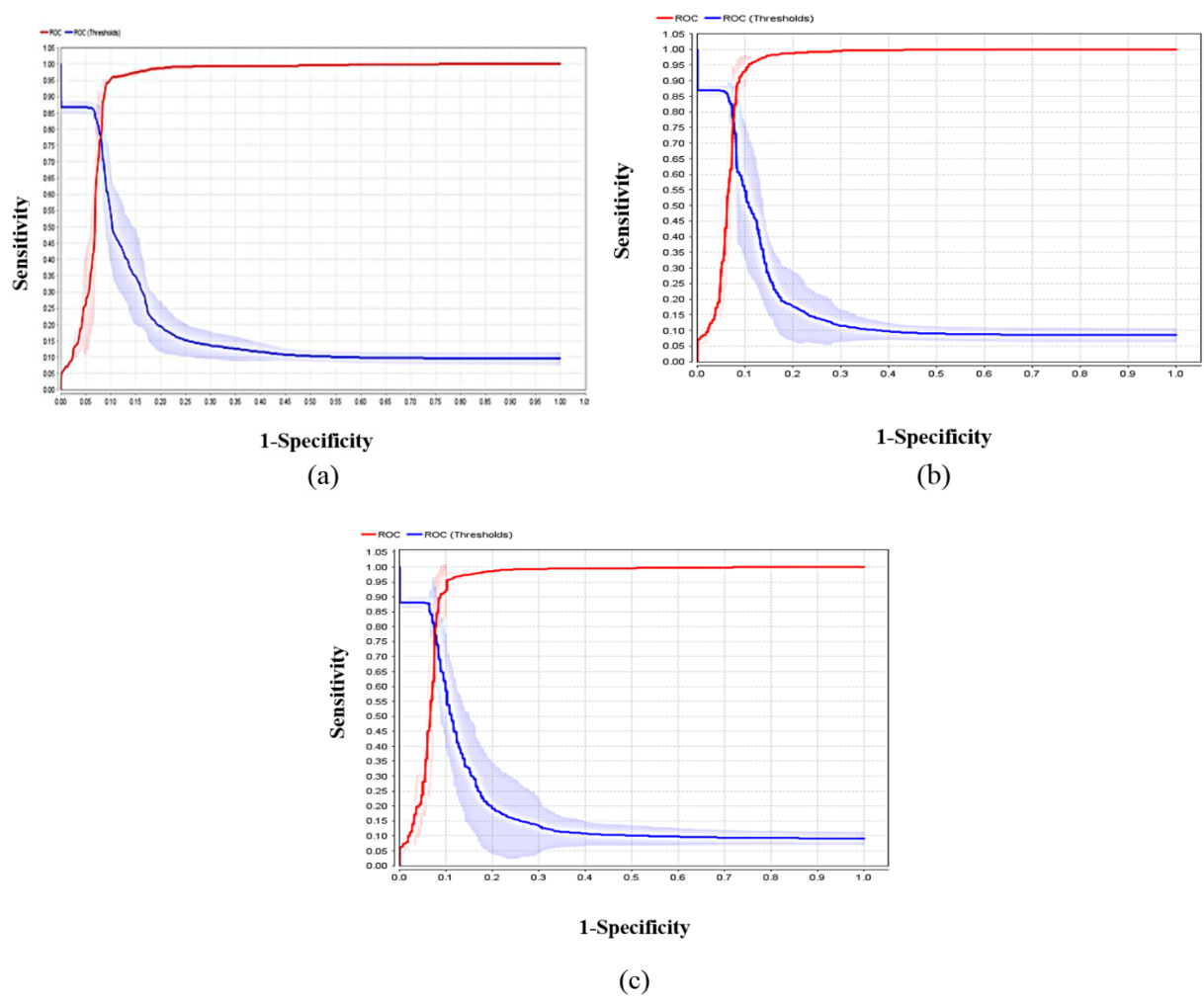

**Figure 10.** (a). The AUC diagram for the NN method through the 5-FCV technique. (b). The AUC diagram for the NN method through the 7-FCV technique. (c). The AUC diagram for the NN method through the 10-FCV technique.

In addition, the FPR and FNR criteria are essential for determining the false rate of diagnosing the disease for clinical centers so that the FPR is more valuable than the FNR for identifying more risks. The FPR value is achieved as zero, while the FNR value is gained as 0.18 using the proposed

---

[1]https://docs.rapidminer.com/latest/studio/operators/modeling/predictive/neural_nets/deep_learning.html

FCM-DNN method. Furthermore, utilizing the DNN method, the value of the FPR is gained as 0.42, and the FNR value is obtained as 0.32. Moreover, applying the NN method, the FPR value is calculated as 12.21, while the FNR value is obtained as 3.56. As a result, the FCM-DNN method has a lower false rate than the NN and DNN classification methods.

As a significant result, there is a crucial criterion for evaluating the classification models, namely AUC, which indicates the accuracy of the level below the receiver operating characteristic (ROC) curve. Based on the 5-FCV, 7-FCV and 10-FCV techniques, the ROC diagram for the NN, DNN and FCM-DNN models are shown in Figures 10(a)–(c), 11(a),(b) and 12, respectively.

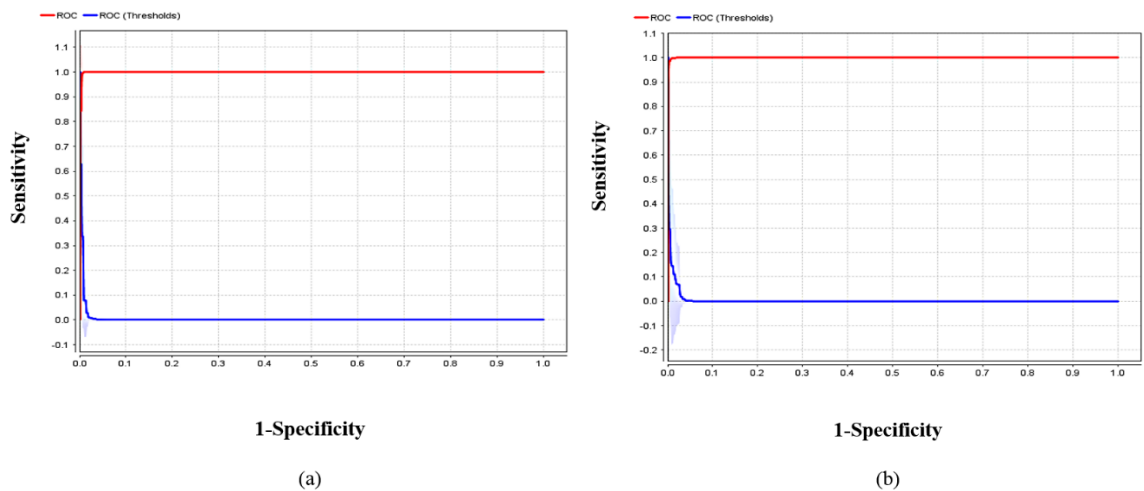

(a)                                    (b)

**Figure 11.** (a). The AUC diagram for the DNN method through the 5-FCV and 10-FCV techniques. (b). The AUC diagram for the DNN method through the 7-FCV technique.

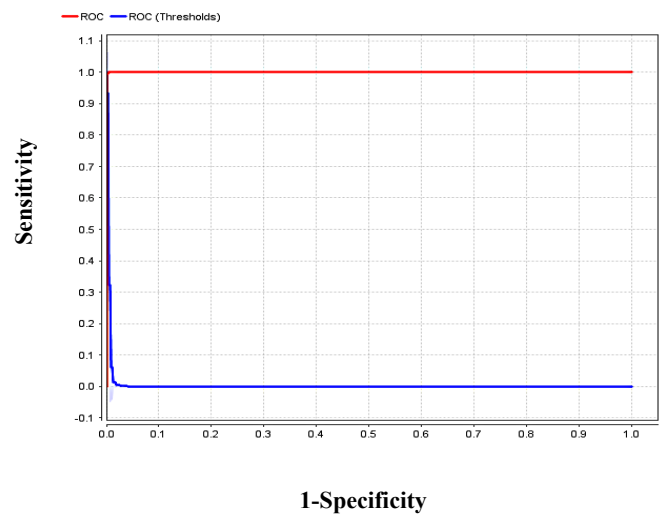

**Figure 12.** The AUC diagram for the FCM-DNN method through the 10-FCV, 7-FCV, and 5-FCV techniques.

According to Figure 10(a), the AUC value for the NN method through the 5-FCV technique is obtained as 93.4%. Moreover, the AUC values of the NN method are achieved as 93.6% and 93.3% through the 7-FCV and 10-FCV techniques, as shown in Figures 10(b),(c), respectively. The AUC value for the DNN method based on the 10-FCV and 5-FCV techniques is gained as 99.9%, as illustrated in Figure 11. The AUC value for the DNN method based on the 7-FCV technique is computed as 100%, as shown in Figure 11(b). Ultimately, according Figure 12, the AUC value for the FCM-DNN method through the 10-FCV, 7-FCV and 5-FCV techniques is obtained as 100%.

As a result, the FCM-DNN model has the best AUC value compared to the NN and DNN models using the 10-FCV, 7-FCV and 5-FCV techniques.

## 4.  Discussion

Recent advances in artificial intelligence methods using image processing for CAD diagnosis have attracted more researchers to the subject. Automatic diagnosis of CAD among sick and healthy images can be a crucial step for medical exegesis utilizing artificial intelligence methods. Deep learning method is the most common method for image processing. In this paper, the 8-layer deep learning model combined with fuzzy C-means clustering has been used for CAD diagnosis. Meanwhile, neural network and deep neural network methods have been implemented and evaluated. In fact, three methods were employed on the CMRI dataset for the first time. Moreover, 10-fold cross-validation, 7-fold cross-validation, and 5-fold cross-validation techniques have been utilized to evaluate the models. The experimental results have demonstrated that the proposed deep learning model improves the automatic diagnosis of CAD in terms of accuracy, precision, sensitivity, specificity, F1-score, false positive rate, false negative rate, and AUC value.

The performance of the proposed models is compared based on various criteria in Figure 13.

According to Figure 13, the FCM-DNN method has the best performance compared to the NN and DNN methods in terms of the evaluation criteria. Therefore, diagnosis of CAD is guaranteed using the FCM-DNN method.

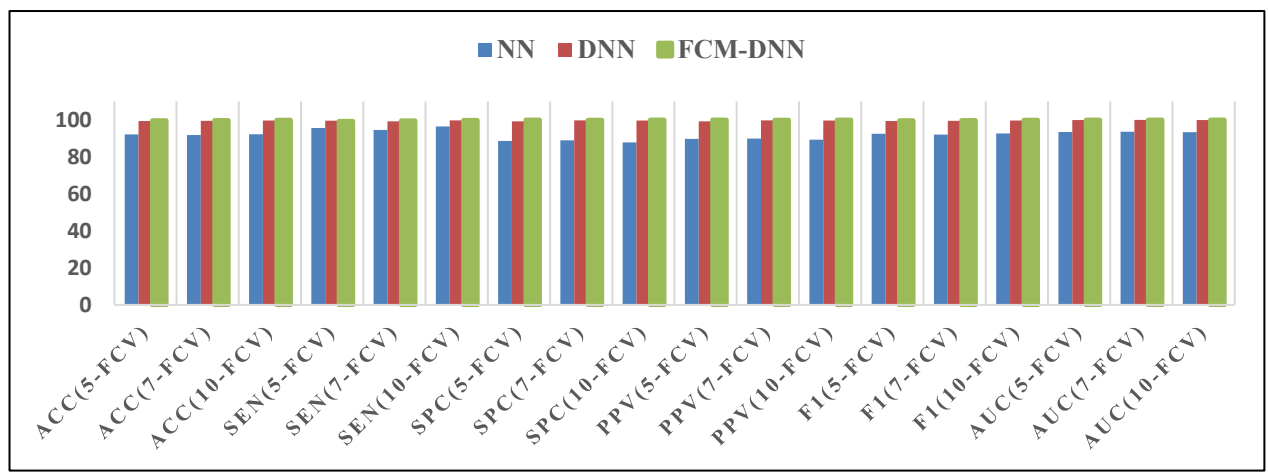

**Figure 13.** A comparison between the NN, DNN, and FCM-DNN methods in terms of the evaluation criteria.

Previous researches as well as the present study are compared in Table 8 regarding the accuracy of CAD diagnosis, methods, and datasets.

**Table 8.** CAD diagnosis using artificial intelligence methods.

| Authors | ACC (%) | Methods | Dataset |
|---|---|---|---|
| Verma et al. [42] | 88.4 | MLR | 335 samples |
| Arabasadi et al. [31] | 93.85 | NN-GA (10-FCV) | 303 samples |
| Alizadehsani et al. [32] | 96.4 | SVM (10-FCV) | 500 samples |
| Miaoa and Miaoa. [34] | 83.67 | DNN | 303 samples |
| Abdar et al. [33] | 93.08 | N2Genetic-NuSVM (10-FCV) | 303 samples |
| Hassannataj et al. [17] | 91.47 | RTs(10-FCV) | 303 samples |
| Ghiasi et al. [41] | 100 | CART | 303 samples |
| Idris et al. [43] | 94.5 | NN with embedded decision tree features | 5100 samples |
| Velusamy and Ramasamy, [44] | 98.97 | weighted-average voting | 303 samples |
| Acharia et al. [39] | 98.5 | KNN (10-FCV) | 52091S ECG signals |
| Tan et al. [38] | 99.85 | LSTM-CNN | 38120 ECG signals |
| Hamersvelt et al. [35] | 71.7 | CNN | Angiography CT images with 126 patients |
| Acharya et al. [40] | 98.97 | CNN (10-FCV) | 140000 ECG signals |
| In this paper | 99.91 | FCM-DNN | 4965 CMRI |

According to Table 8, previous studies have been carried out on three types of datasets, including numerical data, ECG signal data, and CT angiographic data to diagnose CAD. For the first time, we have applied the AI methods on the CMRI image set, and among the NN, DNN, and FCM-DNN methods, we have achieved the highest accuracy of 99.91% for CAD diagnosis on 4965 images using the hybrid FCM-DNN method.

## 5. Conclusions and future work

Coronary artery disease, also known as coronary artery stenosis, is the most common disease in middle-aged and older people. Heart disease [65] is occurred by the accumulation of platelets in the arteries. Following this event, blood flow is clogged, leading to heart failure. The most popular tool for diagnosing CAD disease is angiography, which has side effects and high costs [66].

In recent years, many studies have been conducted to develop artificial intelligence-based methods and replace them with angiography. Hence, in this paper, the NN, DNN, and FCM-DNN methods were applied for CAD diagnosis on the CMRI dataset. The main purpose was to analyze the CMRI dataset in two different approaches using the standard NN, DNN, and FCM-DNN methods. In the first approach, the labeled dataset was applied for the NN and DNN modeling, while in the second approach, the unlabeled dataset was clustered and used for the FCM-DNN modeling.

The results demonstrated that the proposed FCM-DNN method has the best accuracy rate of 99.91% and the least false rate compared to the NN and DNN methods. As a significant achievement, no studies have been carried out for CAD diagnosis on the CMRI dataset so far. As future work, we will study convolutional neural network and auto-encoder neural network algorithms on the CMRI dataset to diagnose CAD.

## Acknowledgments

This article has been funded by Dana Pecutan FTSM (PP-FTSM-2022), Universiti Kebangsaan Malaysia.

## Conflict of interest

The authors declare no conflicts of interest in this paper.

## Availability of data and material

Data and material are available from the corresponding author upon request.

## References

1.  F. Jiang, Y. Jiang, H. Zhi, Y. Dong, H. Li, S. Ma, et al., Artificial intelligence in healthcare: past, present and future, *Stroke Vasc. Neurol.*, **2** (2017). https://doi.org/10.1136/svn-2017-000101

2.  M. D. McCradden, E. A. Stephenson, J. A. Anderson, Clinical research underlies ethical integration of healthcare artificial intelligence, *Nat. Med.*, **26** (2020), 1325–1326. https://doi.org/10.1038/s41591-020-1035-9

3.  K. H. Yu, A. L. Beam, I. S. Kohane, Artificial intelligence in healthcare, *Nat. Biomed. Eng.*, **2** (2018), 719–731. https://doi.org/10.1038/s41551-018-0305-z

4.  O. Asan, A. E. Bayrak, A. Choudhury, Artificial intelligence and human trust in healthcare: focus on clinicians, *J. Med. Internet Res.*, **22** (2020), e15154. https://doi.org/10.2196/15154

5.  D. Shen, G. Wu, H. I. Suk, Deep learning in medical image analysis, *Annu. Rev. Biomed. Eng.*, **19** (2017), 221–248. https://doi.org/10.1146/annurev-bioeng-071516-044442

6.  G. Litjens, T. Kooi, B. E. Bejnordi, A. A. A. Setio, F. Ciompi, M. Ghafoorian, et al., A survey on deep learning in medical image analysis, *Med. Image Anal.*, **42** (2017), 60–88. https://doi.org/10.1016/j.media.2017.07.005

7.  M. I. Razzak, S. Naz, A. Zaib, Deep learning for medical image processing: Overview, challenges and the future, *Classif. BioApps*, (2018), 323–350.

8.  J. H. Thrall, D. Fessell, P. V. Pandharipande, Rethinking the approach to artificial intelligence for medical image analysis: the case for precision diagnosis, *J. Am. Coll. Radiol.*, **18** (2021), 174–179. https://doi.org/10.1016/j.jacr.2020.07.010

9.  Y. Zhang, Z. Wang, J. Zhang, C. Wang, Y. Wang, H. Chen, et al., Deep learning model for classifying endometrial lesions, *J. Transl. Med.*, **19** (2021), 1–13. https://doi.org/10.1186/s12967-020-02660-x

10. C. Zheng, L. Chen, J. Jian, J. Li, Z. Gao, Efficacy evaluation of interventional therapy for primary liver cancer using magnetic resonance imaging and CT scanning under deep learning and treatment of vasovagal reflex, *J. Supercomput.*, **77** (2021), 7535–7548. https://doi.org/10.1007/s11227-020-03539-w

11. G. A. Roth, C. Johnson, A. Abajobir, F. Abd-Allah, S. F. Abera, G. Abyu, et al., Global, regional, and national burden of cardiovascular diseases for 10 causes, 1990 to 2015, *J. Am. Coll. Cardiol.*, **70** (2017), 1–25.

12. K. H. Miao, J. H. Miao, Coronary heart disease diagnosis using deep neural networks, *Int. J. Adv. Comput. Sci. Appl.*, **9** (2018), 1–8. https://doi.org/10.14569/IJACSA.2018.091001

13. A. Gupta, H. S. Arora, R. Kumar, B. Raman, DMHZ: a decision support system based on machine computational design for heart disease diagnosis using z-alizadeh sani dataset, in *2021 International Conference on Information Networking (ICOIN)*, (2021), 818–823. https://doi.org/10.1109/ICOIN50884.2021.9333884

14. A. D. Villa, E. Sammut, A. Nair, R. Rajani, R. Bonamini, A. Chiribiri, Coronary artery anomalies overview: the normal and the abnormal, *World J. Radiol.*, **8** (2016), 537. https://doi.org/10.4329/wjr.v8.i6.537

15. R. Alizadehsani, M. Abdar, M. Roshanzamir, A. Khosravi, P. M. Kebria, F. Khozeimeh, et al., Machine learning-based coronary artery disease diagnosis: A comprehensive review, *Comput. Biol. Med.*, **111** (2019), 103346. https://doi.org/10.1016/j.compbiomed.2019.103346

16. T. M. Williamson, C. Moran, A. McLennan, S. Seidel, P. P. Ma, M. L. Koerner, T. S. Campbell, Promoting adherence to physical activity among individuals with cardiovascular disease using behavioral counseling: A theory and research-based primer for health care professionals, *Prog. Cardiovasc. Dis.*, (2020). https://doi.org/10.1016/j.pcad.2020.12.007

17. J. H. Joloudari, E. H. Joloudari, H. Saadatfar, M. Ghasemigol, S. M. Razavi, A. Mosavi, et al., Coronary artery disease diagnosis; ranking the significant features using a random trees model, *Int. J. Environ. Res. Public Health*, **17** (2020), 731. https://doi.org/10.3390/ijerph17030731

18. M. V. Dyke, S. Greer, E. Odom, L. Schieb, A. Vaughan, M. Kramer, et al., Heart disease death rates among blacks and whites aged ≥ 35 years–United States, 1968–2015, *MMWR Surveillance Summaries*, **67** (2018), 1. https://doi.org/10.15585/mmwr.ss6705a1

19. D. Mozaffarian, E. J. Benjamin, A. S. Go, D. K. Arnett, M. J. Blaha, M. Cushman, et al., Heart disease and stroke statistics—2015 update: a report from the American heart association, *Circulation*, **131** (2015), e29–e322.

20. E. J. Benjamin, S. S. Virani, C. W. Callaway, A. M. Chamberlain, A. R. Chang, S. Cheng, et al., Heart disease and stroke statistics—2018 update: a report from the American heart association, *Circulation*, **137** (2018), e67–e492. https://doi.org/10.1161/CIR.0000000000000573

21. H. Larochelle, Y. Bengio, J. Louradour, P. Lamblin, Exploring strategies for training deep neural networks, *J. Mach. Learn. Res.*, **10** (2009).

22. R. O. Bonow, D. L. Mann, D. P. Zipes, P. Libby, *Braunwald's Heart Disease E-Book: A Textbook of Cardiovascular Medicine*, Elsevier Health Sciences, 2011.

23. E.G. Nabel, E. Braunwald, A tale of coronary artery disease and myocardial infarction, *New Engl. J. Med.*, **366** (2012), 54–63. https://doi.org/10.1056/NEJMra1112570

24. İ. Babaoglu, O. Findik, E. Ülker, A comparison of feature selection models utilizing binary particle swarm optimization and genetic algorithm in determining coronary artery disease using support vector machine, *Expert Syst. Appl.*, **37** (2010), 3177–3183. https://doi.org/10.1016/j.eswa.2009.09.064

25. M. Kumar, R. B. Pachori, U. R. Acharya, Characterization of coronary artery disease using flexible analytic wavelet transform applied on ECG signals, *Biomed. Signal Proces. Control*, **31** (2017), 301–308. https://doi.org/10.1016/j.bspc.2016.08.018

26. R. Alizadehsani, J. Habibi, M. J. Hosseini, R. Boghrati, A. Ghandeharioun, B. Bahadorian, et al., Diagnosis of coronary artery disease using data mining techniques based on symptoms and ecg features, *Eur. J. Scientific Res.*, **82** (2012), 542–553.

27. R. Alizadehsani, J. Habibi, M. J. Hosseini, H. Mashayekhi, R. Boghrati, A. Ghandeharioun, et al., A data mining approach for diagnosis of coronary artery disease, *Comput. Meth. Prog. Bio.*, **111** (2013), 52–61. https://doi.org/10.1016/j.cmpb.2013.03.004

28. R. Alizadehsani, J. Habibi, Z. A. Sani, H. Mashayekhi, R. Boghrati, A. Ghandeharioun, et al., Diagnosing coronary artery disease via data mining algorithms by considering laboratory and echocardiography features, *Res. Cardiov. Med.*, **2** (2013), 133. https://doi.org/10.5812/cardiovascmed.10888

29. R. Alizadehsani, M. H. Zangooei, M. J. Hosseini, J. Habibi, A. Khosravi, M. Roshanzamir, et al., Coronary artery disease detection using computational intelligence methods, *Knowledge-Based Syst.*, **109** (2016), 187–197. https://doi.org/10.1016/j.knosys.2016.07.004

30. A. D. Dolatabadi, S. E. Z. Khadem, B. M. Asl, Automated diagnosis of coronary artery disease (CAD) patients using optimized SVM, *Comput. Meth. Prog. Bio.*, **138** (2017), 117–126. https://doi.org/10.1016/j.cmpb.2016.10.011

31. Z. Arabasadi, R. Alizadehsani, M. Roshanzamir, H. Moosaei, A. A. Yarifard, et al., Computer aided decision making for heart disease detection using hybrid neural network-Genetic algorithm, *Comput. Meth. Prog. Bio.*, **141** (2017), 19–26. https://doi.org/10.1016/j.cmpb.2017.01.004

32. R. Alizadehsani, M. J. Hosseini, A. Khosravi, F. Khozeimeh, M. Roshanzamir, N. Sarrafzadegan, et al., Non-invasive detection of coronary artery disease in high-risk patients based on the stenosis prediction of separate coronary arteries, *Comput. Meth. Prog. Bio.*, **162** (2018), 119–127. https://doi.org/10.1016/j.cmpb.2018.05.009

33. M. Abdar, W. Książek, U. R. Acharya, R. S. Tan, V. Makarenkov, P. Pławiak, A new machine learning technique for an accurate diagnosis of coronary artery disease, *Comput. Meth. Prog. Bio.*, **179** (2019), 104992. https://doi.org/10.1016/j.cmpb.2019.104992

34. C. Blake, *UCI Repository of Machine Learning Databases*, 1998. Available from: http://www.ics.uci.edu/~mlearn/MLRepository.html.

35. R. W. Hamersvelt, M. Zreik, M. Voskuil, M. A. Viergever, I. Išgum, T. Leiner, et al., Deep learning analysis of left ventricular myocardium in CT angiographic intermediate-degree coronary stenosis improves the diagnostic accuracy for identification of functionally significant stenosis, *Eur. Rad.*, **29** (2019), 2350–2359. https://doi.org/10.1007/s00330-018-5822-3

36. U. R. Acharya, H. Fujita, O. S. Lih, M. Adam, J. H. Tan, C. K. Chua, Automated detection of coronary artery disease using different durations of ECG segments with convolutional neural network, *Knowledge-Based Syst.*, **132** (2017), 62–71. https://doi.org/10.1016/j.knosys.2017.06.003

37. A. L. Goldberger, L. A. Amaral, L. Glass, J. M. Hausdorff, P. C. Ivanov, R. G. Mark, et al., PhysioBank, PhysioToolkit, and PhysioNet: components of a new research resource for complex physiologic signals, *Circulation*, **101** (2000), e215–e220. https://doi.org/10.1161/01.CIR.101.23.e215

38. J. H. Tan, Y. Hagiwara, W. Pang, I. Lim, S. L. Oh, M. Adam, et al., Application of stacked convolutional and long short-term memory network for accurate identification of CAD ECG signals, *Comput. Biol. Med.*, **94** (2018), 19–26. https://doi.org/10.1016/j.compbiomed.2017.12.023

39. U. R. Acharya, H. Fujita, M. Adam, O. S. Lih, V. K. Sudarshan, T. J. Hong, et al., Automated characterization and classification of coronary artery disease and myocardial infarction by decomposition of ECG signals: A comparative study, *Inform. Sci.*, **377** (2017), 17–29. https://doi.org/10.1016/j.ins.2016.10.013

40. U. R. Acharya, H. Fujita, S. L. Oh, Y. Hagiwara, J. H. Tan, M. Adam, et al., Deep convolutional neural network for the automated diagnosis of congestive heart failure using ECG signals, *Appl. Intell.*, **49** (2019), 16–27. https://doi.org/10.1007/s10489-018-1179-1

41. M. M. Ghiasi, S. Zendehboudi, A. A. Mohsenipour, Decision tree-based diagnosis of coronary artery disease: CART model, *Comput. Meth. Prog. Bio.*, **192** (2020), 105400. https://doi.org/10.1016/j.cmpb.2020.105400

42. L. Verma, S. Srivastava, P. Negi, A hybrid data mining model to predict coronary artery disease cases using non-invasive clinical data, *J. Med. Syst.*, **40** (2016), 178. https://doi.org/10.1007/s10916-016-0536-z

43. N. M. Idris, Y. K. Chiam, K. D. Varathan, W. A. W. Ahmad, K. H. Chee, Y. M. Liew, Feature selection and risk prediction for patients with coronary artery disease using data mining, *Med. Biol. Eng. Comput.*, **58** (2020), 3123–3140. https://doi.org/10.1007/s11517-020-02268-9

44. D,. Velusamy, K. Ramasamy, Ensemble of heterogeneous classifiers for diagnosis and prediction of coronary artery disease with reduced feature subset, *Comput. Meth. Prog. Bio.*, **198** (2020), 105770. https://doi.org/10.1016/j.cmpb.2020.105770

45. I. Goodfellow, D. Warde-Farley, M. Mirza, A. Courville, Y. Bengio, Maxout networks, in *International Conference On Machine Learning*, PMLR, (2013), 1319–1327.

46. J. H. Joloudari, M. Haderbadi, A. Mashmool, M. GhasemiGol, S. S. Band, A. Mosavi, Early detection of the advanced persistent threat attack using performance analysis of deep learning, *IEEE Access*, **8** (2020), 186125–186137. https://doi.org/10.1109/ACCESS.2020.3029202

47. Y. Ito, Approximation of functions on a compact set by finite sums of a sigmoid function without scaling, *Neural Networks*, **4** (1991), 817–826. https://doi.org/10.1016/0893-6080(91)90060-I

48. N. Hassan, N. Akamatsu, A new approach for contrast enhancement using sigmoid function, *Inter Arab J. Inf. Techn.*, **1** (2004).

49. X. Li, X. Zhang, W. Huang, Q. Wang, Truncation cross entropy loss for remote sensing image captioning, *IEEE Transactions Geosci. Remote Sens.*, **59** (2020), 5246–5257. https://doi.org/10.1109/TGRS.2020.3010106

50. C. Otto, D. Wang, A. K. Jain, Clustering millions of faces by identity, *IEEE Transactions Pattern Anal. Mach. Intell.*, **40** (2017), 289–303. https://doi.org/10.1109/TPAMI.2017.2679100

51. E. H. Ruspini, A new approach to clustering, *Inform. Control*, **15** (1969), 22–32. https://doi.org/10.1016/S0019-9958(69)90591-9

52. R. O. Duda, P. E. Hart, *Hart PE Pattern Classification And Scene Analysis*, New York: Wiley, 1973.

53. R. Veloso, F. Portela, M. F. Santos, A. Silva, F. Rua, A. Abelha, et al., A clustering approach for predicting readmissions in intensive medicine, *Procedia Technol.*, **16** (2014), 1307–1316. https://doi.org/10.1016/j.protcy.2014.10.147

54. H. S. Park, C. H. Jun, A simple and fast algorithm for K-medoids clustering, *Expert Syst. Appl.*, **36** (2009), 3336–3341. https://doi.org/10.1016/j.eswa.2008.01.039

55. R. O. Duda, P. E. Hart, *Pattern Classification And Scene Analysis*, Wiley New York, 1973.

56. J. C. Dunn, Well-separated clusters and optimal fuzzy partitions, *J. Cybern.*, **4** (1974), 95–104. https://doi.org/10.1080/01969727408546059

57. J. C. Bezdek, Objective function clustering, in *Pattern Recognition With Fuzzy Objective Function Algorithms*, Springer, (1981), 43−93. https://doi.org/10.1007/978-1-4757-0450-1_3

58. M. S. Yang, A survey of fuzzy clustering, *Math. Comput. Model.*, **18** (1993), 1−16. https://doi.org/10.1016/0895-7177(93)90202-A

59. G. Govaert, M. Nadif, Clustering with block mixture models, *Pattern Recogn.*, **36** (2003), 463−473. https://doi.org/10.1016/S0031-3203(02)00074-2

60. S. Bandyopadhyay, U. Maulik, A. Mukhopadhyay, Multiobjective genetic clustering for pixel classification in remote sensing imagery, *IEEE Trans. Geosci. Remote Sens.*, **45** (2007), 1506−1511. https://doi.org/10.1109/TGRS.2007.892604

61. R. Xu, D. Wunsch, Survey of clustering algorithms, *IEEE Trans. Neural Networks*, **16** (2005), 645−678. https://doi.org/10.1109/TNN.2005.845141

62. J. H. Joloudari, H. Saadatfar, A. Dehzangi, S. Shamshirband, Computer-aided decision-making for predicting liver disease using PSO-based optimized SVM with feature selection, *Inform. Medicine Unlocked*, **17** (2019), 100255. https://doi.org/10.1016/j.imu.2019.100255

63. M. Abdar, M. Zomorodi-Moghadam, R. Das, I. H. Ting, Performance analysis of classification algorithms on early detection of liver disease, *Expert Syst. Appl.*, **67** (2017), 239−251. https://doi.org/10.1016/j.eswa.2016.08.065

64. C. H. Weng, C. K. Huang, R. P. Han, Disease prediction with different types of neural network classifiers, *Telemat. Inform.*, **33** (2016), 277−292. https://doi.org/10.1016/j.tele.2015.08.006

65. M. Diwakar, A. Tripathi, K. Joshi, M. Memoria, P. Singh, Latest trends on heart disease prediction using machine learning and image fusion, *Materials Today: Proceed.*, **37** (2021), 3213−3218. https://doi.org/10.1016/j.matpr.2020.09.078

66. J. H. Moon, W. C. Cha, M. J. Chung, K. S. Lee, B. H. Cho, J. H. Choi, Automatic stenosis recognition from coronary angiography using convolutional neural networks, *Comput. Meth. Prog. Bio.*, **198** (2021), 105819. https://doi.org/10.1016/j.cmpb.2020.105819

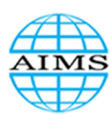 AIMS Press